# Circuit Level Modeling of Extra Combinational Delays in SRAM FPGAs Due to Transient Ionizing Radiation

Mostafa Darvishi[1], *Student Member, IEEE*, Yves Audet[1], *Member, IEEE*, Yves Blaquière[2], *Member, IEEE*, and Claude Thibeault[3], *Senior Member, IEEE*

[1]Electrical Engineering Department, École Polytechnique de Montreal, Montreal, QC, Canada

[2]Computer Science Department, Université du Québec à Montréal, Montreal, QC, Canada

[3]Electrical Engineering Department, École de Technologie Supérieure, Montreal, QC, Canada

*Abstract*

This paper presents a novel circuit level model that explains and confirms the extra combinational delays in a SRAM-FPGA (Virtex-5) due to radiation, which matches the experimental results by proton irradiation at TRIUMF.

**Presenter:**
Mostafa Darvishi,
Ph.D. Student
École Polytechnique de Montreal
Electrical Engineering Department
CP. 6079, succ. Centre-Ville,
 Montréal, QC, Canada, H3C 3A7
tel 514 340-4711
fax 514 340-4147
Email: mostafa.darvishi@polymtl.ca

## I. INTRODUCTION

SRAM-based Field Programmable Gate Arrays (FPGAs) are semiconductor devices that are based around an array of configurable logic blocks (CLBs) connected via a hierarchy of configurable interconnects. FPGAs have become the preferred common solution to implement digital systems targeting different applications. The SRAM-based FPGA comprises some I/O blocks, memory modules, logic blocks and routing resources controlled by SRAM cells, called configuration bits [1]. The sensitivity to radiation of SRAM-based FPGAs has been studied over the years [2, 3]. The first report on extra combinational delays due to transient ionizing radiations was presented in [4] where the existence of extra delays due to Single-Event-Upsets (SEUs) induced by proton radiation was experimentally observed.

The main contribution of this paper is the validation of the root cause of Observed Delay Changes (ODCs) on SRAM-based FPGA through circuit level simulations of the internal circuitry of Configurable Logic Blocks (CLBs) [5-7] and their interconnections.

This summary presents a novel circuit model created to understand and simulate the source of extra combinational delays experimentally observed that are ranging from 40 ps to as much as 422 ps [4]. To our knowledge, the proposed model and methodology represents the first work ever on the simulation of extra combinational delays due to SEU occurring in FPGAs. The model is accurate enough to obtain close correlation with the experimental results. The proposed methodology can also be used to predict the probable delay values due to radiation in any design implemented on FPGA.

This paper is structured as follows. Some background information regarding the previous work is presented in Section II. Section III introduces the FPGA circuit level model for ODC root cause validation including circuit level model and model configuration tuning, respectively. Typical circuit-level configurations that could induce ODC in SRAM FPGA are presented in Section IV. Comparison between simulation results with ones experimentally observed by proton irradiation is discussed in Section V, and we conclude in Section VI.

## II. BACKGROUND OF THE PREVIOUS WORK

Configuration memory cells in SRAM-based FPGA are sensitive to radiation that causes a bit flip of the stored values. These SRAMs are mainly used to configure interconnects and look-up tables. The two impacts of a bit-flip on configuration bits related to interconnections are open (namely a disappearing link between two nodes) and short (usually defined an undesired connection between two routed signals) faults. While SEU can modify logic behavior in SRAM-based FPGA, it was conjectured in [4] that delays could be induced by a different type of short, between a routed signal and an unused wire.

Fig. 1 illustrates the experimental setup that was utilized at the TRIUMF laboratory to demonstrate these induced extra delays in SRAM-FPGA. Extra combinational delays were observed while the board was bombarded by protons (35.4 MeV, 50 MeV, 57.7 MeV, 63 MeV and 105 MeV) for several runs. The XilinxVirtex-5 FPGA was used to implement two ring oscillators (ROs) made of inverters operating at similar frequencies. The output of each ring oscillator was connected to one external inverter (7404). The outputs of the two 7404 inverters were shorted by a 5.1 kΩ resistor while one inverter output is monitored by a spectrum analyzer. The resistive shorted outputs provide a signal with a frequency spectrum containing the difference frequency between two RO frequencies ($F_2 - F_1$). This difference is mainly due to the parameter variation in fabrication process and slight difference in the oscillator's routing. The measurement of the difference ($F_2 - F_1$) instead of individually measuring $F_1$ and $F_2$ led to a better precision. The ring oscillators were adjusted to the length of 1799 inverters creating $F_1 \approx F_2 \approx 1.25$ MHz and a frequency difference of about 12.4 kHz.

Consequently, a set of 48 experiments were performed in [4] with the proton source bombarding the top side of the FPGA. Each delay measurement was stopped when one RO broke and 23 of those experiments came with one or cumulative ODCs. The delay change could produce either a reduction or an increase of the measured frequency difference depending on which of the ring oscillators was affected.

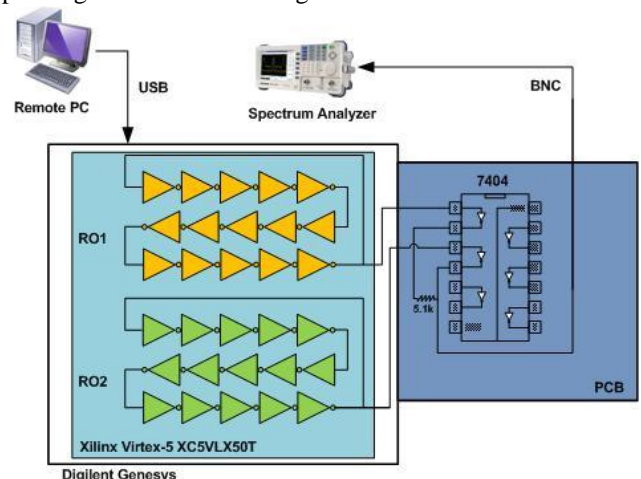

Fig. 1. Experimental setup at TRIUMF.

## III. FPGA CIRCUIT-LEVEL MODEL FOR ODC ROOT CAUSE VALIDATION

### A. Circuit Level Model

One contribution of this paper is to present a circuit level model of the FPGA that takes into account the CLB along with their interconnection modules in order to simulate SEU induced delays. The Virtex-5 is based on an array of Configurable Logic Blocks with 4 slices each [1]. The circuit is modeled as a two dimensional array comprising slices, programmable interconnection points (PIP) and switch boxes (SB) interconnected by a network of horizontal and vertical routing wires as shown in Fig. 2. Xilinx does not formally provide details on internal Virtex-5 FPGA circuitries. However according to [5, 6], PIP and SB are made of one nMOS pass transistor while a slice includes a Configurable

Logic Element (CLE) coupled to the general interconnect structure via input multiplexers (IMUX). The CLE is comprised of a look-up table connected to a multiplexer, and the IMUX is composed of an 8:1 multiplexer connected to a regenerator circuit.

Fig. 2 presents the top level view of two adjacent CLBs, where two different configurations of slice-to-slice interconnection are shown as examples. The first configuration, path A-to-B, is introduced to simulate the behavior of direct slice-to-slice link of the ring oscillator (RO) between two adjacent CLBs. The second configuration, path C-to-A, represents the other possible interconnection between two slices in a same CLB. Both configurations are reported by the Xilinx FPGA Editor tool [8]. In the first configuration, the CLB-to-CLB interconnection length, $L_{CC}$, is longer than the Slice-to-SB interconnection length, $L_{SS}$, in the second configuration. These configurations are introduced as models for ring oscillators implemented on FPGA enabling the prediction of the probable ODCs.

*B. Model Configuration Tuning*

The circuit models employed to simulate both interconnect configurations of the RO implementation used in the experiments are detailed in Fig. 3(a) (path A-to-B) and Fig. 3(b) (path C-to-A), respectively. A signal shaping filter comprised of four inverters generates a realistic pulse signal waveform. According to Fig. 3(a), any interconnection between two slices located in two adjacent CLBs has to pass through two switch boxes with an interconnection length of $L_{CC}$. In Fig. 3(b), the interconnection between two slices located in a same CLB passes through a switch box with two interconnections of length $L_{SS}$. The switch box is comprised of an array of pass transistors and very short interconnections shown as $L_{PP}$ in Fig. 3(b). Our simulations show that the effect of $L_{PP}$ on propagation delays is negligible compared to the one of a pass transistor along the path.

Fig. 4 presents the propagation delay results from Slice-to-SB (node C to node A) and Slice-to-Slice (node A to node B) as a function of the interconnection length to adjust $L_{SS}$ and $L_{CC}$ in the first and second configurations. We found that an interconnection length of $L_{SS}$ = 1.74 µm in the second configuration (node C to node A) matches the inverter and net delay of 138 ps that was extracted by Xilinx STA-TRACE. The same procedure was performed to match the value of LCC in the first configuration (node A to node B) and the corresponding value amounts to 7.35 µm, which matches the net delay of 484 ps. Our simulation results showed that the effect of $L_{PP}$ on the delay is negligible compared to PIP's effect, so its value was neglected. As shown in the following section, the adjusted lengths $L_{CC}$ and $L_{SS}$ and our circuit models provided sufficient accuracy to reproduce the ODCs observed experimentally.

IV. CIRCUIT LEVEL CONFIGURATIONS INDUCING OBSERVED DELAY CHANGES (ODCS)

A SEU in SRAM-based FPGA can affect a SRAM-cell by creating a short, an open or a modification in logic behavior.

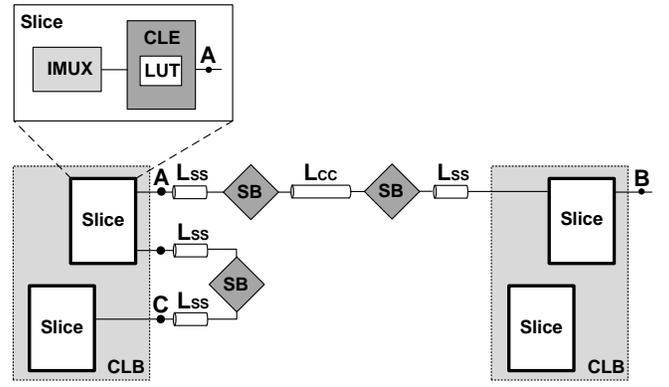

Fig. 2. Model of a two configurations of slice to slice interconnection in Virtex-5.

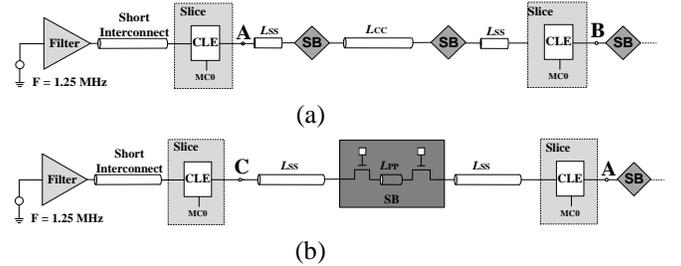

Fig. 3. Structure of slice-to-slice interconnection a) between two adjacent CLBs, b) in a single CLB.

Indeed, it is assumed in this paper as in [4]) that the experimentally observed delay changed is caused by an SEU that increases the interconnect load parasitic capacitance, which increases the routing delay.
An SEU affecting an SRAM-cell controlling a PIP (Programmable Interconnection Point) could create a short, for example, between a vertical line and the main horizontal routing line, as shown in the simplified schematic view of Fig.5 (a). According to our simulation results, the main contributor to the delay is the PIP pass-transistor that is turned 'on' and increases the parasitic capacitance by connecting an undesired vertical unused interconnection to the main routing path.

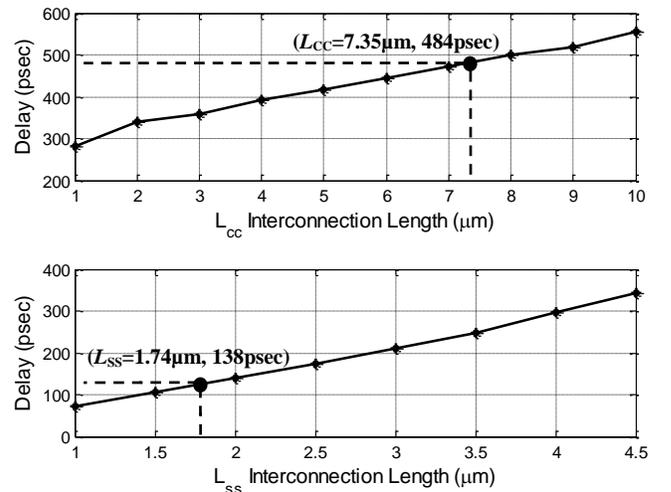

Fig. 4. Delay variation as a function of short interconnection length for the configurations used to define $L_{CC}$ in Fig. 4(a) and $L_{SS}$ in Fig. 4(b).

While Fig. 5 (a) is an example of a single interconnect parasitic (1-SEU case), Fig. 5 (b) shows that cumulative SEUs can create larger parasitic load than the 1-SEU case on the main routing path. In Fig. 5(b), it is assumed that a primary SEU affected the configuration bit of the SRAM-cell and turned on the corresponding PIP pass transistor and made a permanent connection between one horizontal and one vertical interconnect that are not yet connected to the main routing path. The extra capacitance is added on the main routing path when another SEU flips the configuration bit of the SRAM-cell that connects the two former parasitic lines to the main routing path and therefore creates an extra parasitic delay.

Fig. 6 illustrates a 3-SEU case where an even larger combinational delay is created by a sequence of three consecutive SEUs. The first two SEUs enabled $PIP_B$ and $PIP_C$, and then another SEU activated $PIP_A$ to create a combinational delay larger than the one observed in the 2-SEU case.

It is noticeable that the presented structures can be applied for both configurations introduced in Fig. 2.

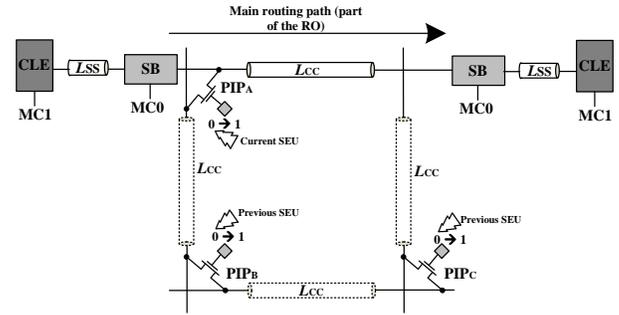

Fig. 6. Effect of SEUs connecting three unwanted interconnects, two vertical and one horizontal.

In our convention, for the configuration presented in Fig. 4(b), the (1) case was simulated, which means 1 SEU has shorted a parasitic interconnect to the main routing path while its length is 1 $L_{SS}$. The (1), (2) and (4) cases were simulated for the configuration presented in Fig. 4(a), while an SEU has connected a parasitic interconnect to the main routing path with the lengths of 1 $L_{CC}$, 2 $L_{CC}$ or 4 $L_{CC}$, respectively. Notice that regarding the probable interconnection lengths in Virtex-5, direct CLB-CLB connections in Virtex-5 FPGA can be 1 $L_{CC}$, 2 $L_{CC}$ or 4 $L_{CC}$ [7, 9], as shown in Fig. 7. More scenarios have been simulated for the configuration of Fig. 3(a) that includes 2-SEU and 3-SEU cases. The nomenclatures of (1,1), (1,2), (1,4), (2,1), (2,2), (2,4), (4,1), (4,2) and (4,4) are defined while the main routing path is affected by 2 SEUs. For instance, the case (1,1) identifies a cumulative case where two parasitic interconnects with the length of 1 $L_{CC}$ due to two consecutive SEUs are connected to the main routing path as shown in Fig. 5(b). Also, the case (1,2) implies two parasitic interconnects with the length of 1 $L_{CC}$ and 2 $L_{CC}$ respectively connected to the main routing path. The case (4,4) represents two parasitic interconnects both with the length of 4 $L_{CC}$ linked to the main routing path.

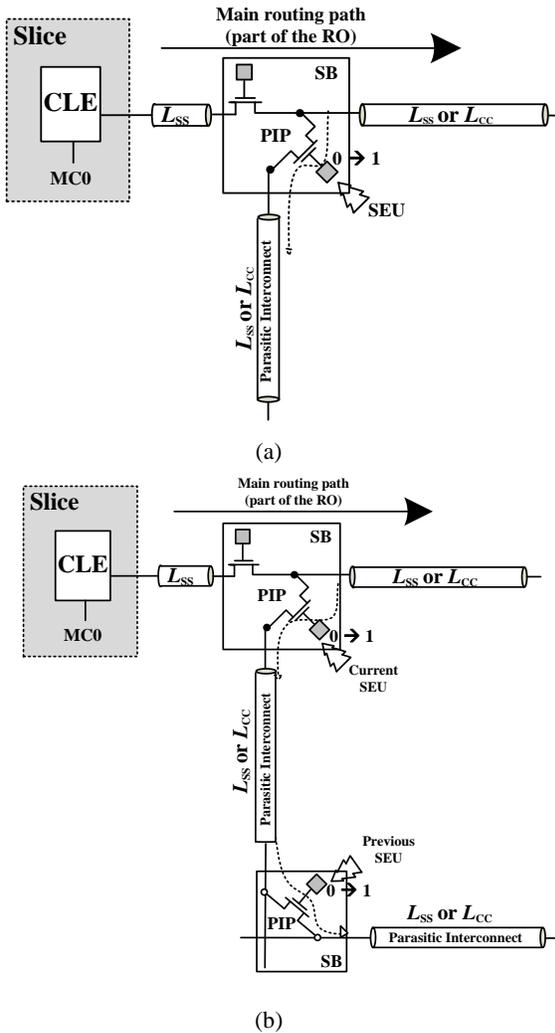

Fig. 5. Effect of an SEU on a Programmable Interconnection Point (PIP) in SB, adding a combinational delay: a) 1 ODC case (1 SEU), b) 2 ODC case (2 SEUs).

## V. COMPARISON OF SIMULATION RESULTS WITH EXPERIMENTAL RESULTS BY PROTON IRRADIATION

Circuit-level simulations of ODCs were performed for the three proposed configurations (Fig. 5 and 6). The results closely correlate experimental results obtained at TRIUMF, as shown in Table 1. Notice also that the index '2nd' stated in Table 1 corresponds to the second configuration shown in Fig. 3(b) that connects a parasitic interconnect to the main routing path while its length could be 1 $L_{SS}$. Circuit-level simulations of ODCs were performed for the three proposed scenarios presented in section IV and illustrated in Fig. 5, 6 for a total of three different scenarios. The simulated delays correlate fairly well with the ODCs measured at TRIUMF.

For the each case of ODC = 2, a combination of simulated cases for ODC = 1 is added in order to match the delay. It is worth mentioning that only simulated delay results that closely match the experimental results by proton irradiation at TRIUMF are provided. Further configuration cases are being investigated to cover all experimental results.

Recall that 1 $L_{CC}$ represents the unit length extracted when tuning our model for the first configuration in Section III,

equal to 7.35 μm. The three introduced scenarios in Section IV due to Extra Parasitic Interconnects were simulated according to the possible interconnection lengths defined in Fig. 7.

Further investigation is under way to refine our model in order to closely match all our measurement results. Long delays obtained from multiple ODCs (more than 2) will be dissected to ease the matching process with simulations and understand the effect of multiple delay occurrences. An updated circuit model with more results will be presented at the time of the conference.

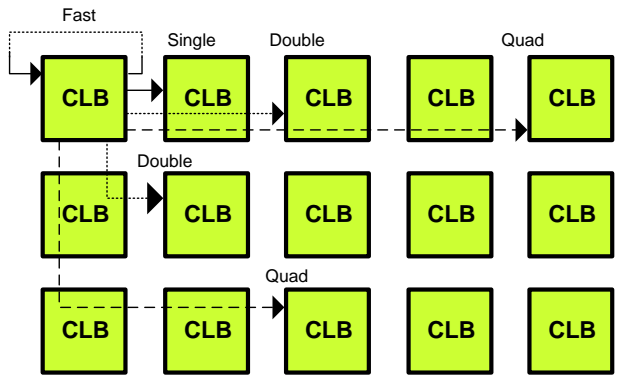

Fig. 7. Examples of various interconnection lengths in Virtex-5 FPGA [7].

TABLE 1
Experimental vs. Simulation Results

| (MEAS: Experimental results observed at TRIUMF, SIM: Simulation results, Model: Configuration Model. | | | | | |
|---|---|---|---|---|---|
| DELAY (ps) | | | | | |
| ODC = 1 | | | ODC = 2 | | |
| MEAS [±6 ps] | SIM | Model | MEAS [±6 ps] | SIM | Model |
| 38 | 39.21 | $(1)_{Lss}$ | 88 | | |
| 49 50 | | | 98 98 | 96 | (1)+(2) |
| 62 | 64.8 | (1,2) | 110 112 | 112 | (2)+(2) |
| | | | 124 | 123 | (2)+(2,1) |

## VI. CONCLUSION

This paper presented results supporting the assumption that extra combinational delays in SRAM FPGAs due to radiations are caused by bit flip of SRAM-cells configuring FPGA interconnection points and switch boxes and adding parasitic capacitance. We proposed a novel circuit level model that has successfully been used to simulate the experimental results obtained with a pair of ring oscillators. Our simulation results closely correlated with those observed at TRIUMF and can describe different scenarios creating delay change in critical routing paths. The proposed methodology can be used to predict the delay value of one or multiple ODCs due to radiation in any design implemented in FPGAs.

## ACKNOWLEDGMENT

The authors would like to thank Natural Sciences and Engineering Research Council of Canada (NSERC) and MITACS for their financial support and CMC Microsystems for its tools and technologies used during this project.